\newcommand\p{\partial}
\newcommand\ep{\epsilon}

\newcommand\pr{\prime}
\newcommand\ola{\overleftarrow}
\newcommand\ora{\overrightarrow}

\newcommand{\lan}{\langle}

\newcommand{\ran}{\rangle}

\newcommand{\diracslash}[1]{#1\llap{/\kern2pt}}

\def\bearr{\begin{eqnarray}}

\def\eearr{\end{eqnarray}}

\newcommand{\be}{\begin{equation}}

\newcommand{\ee}{\end{equation}}

\newcommand{\bea}{\begin{eqnarray}}

\newcommand{\eea}{\end{eqnarray}}

\newcommand{\ba}[1]{\begin{array}{#1}}

\newcommand{\ea}{\end{array}}

\documentclass[preprint,secnumarabic,floats,amssymb,nobibnotes,nofootinbib,aps,prc,showpacs,tightenlines]{revtex4}

\usepackage{graphicx}

\usepackage{amsmath}

\usepackage{latexsym}

\usepackage{epsfig}

\usepackage{subfigure}

\begin{document}
\title{Anisotropic non-gaussianity with noncommutative spacetime}
\author{Akhilesh Nautiyal}
\affiliation{Institute of Mathematical Sciences, CIT Campus, Taramani, Chennai-600113,
 India}
\begin{abstract}
We study single field inflation in noncommutative 
spacetime and compute
two-point and three-point correlation functions for the  curvature
perturbation. We find that both power spectrum and bispectrum for 
comoving curvature perturbation are statistically anisotropic and the
bispectrum is also modified by a phase factor depending upon the
noncommutative parameters. The non-linearity parameter $f_{NL}$ is small 
for small statistical anisotropic corrections to the bispectrum coming from 
the  noncommutative geometry and is consistent with the recent PLANCK bounds. 
There is a scale dependence of $f_{NL}$ due 
to the noncommutative spacetime which is different from the standard 
single field inflation models and statistically anisotropic vector field 
inflation models.
Deviations from statistical isotropy of CMB, observed by PLANCK can tightly
constraint the effects due to noncommutative geometry on power spectrum and 
bispectrum. 
\end{abstract}
\maketitle

\section{Introduction}
Inflation \cite{Guth:1980zm} not only  solves  the various puzzles of the Big-Bang
theory, but it also provides seeds for the temperature anisotropy  
 of the cosmic microwave background (CMB) radiation and structures
in the universe. In the standard inflationary scenario, the potential energy
of a scalar field called "inflaton" dominates the energy density of the 
universe and its quantum fluctuations generate perturbations in the metric 
causing small inhomogeneities in the early universe which give rise to CMB 
anisotropy and structures in the universe.

Inflation predicts nearly scale invariant, adiabatic and 
gaussian perturbations. The first two are in excellent agreement with the 
observations of CMB anisotropy and polarization by COBE \cite{Smoot:1992td}, 
WMAP \cite{Spergel:2003cb,Komatsu:2010fb,Hinshaw:2012fq} and other
 ground based and satellite based experiments, but the test of gaussian 
statistics of the perturbations is controversial and is the major goal
of ongoing and future observations like PLANCK \cite{:2006uk},
 CMBPOL \cite{Baumann:2008aq} and Euclid satellite \cite{Amendola:2012ys}.
Recently released PLANCK data has tightened the bounds on non-gaussianity 
\cite{Ade:2013ydc}.

The non-gaussianity in CMB can be primordial or can be generated due to 
secondary sources (see \cite{Komatsu:2002db} for detailed review). 
The primordial non-gaussianity arises due to the interaction terms in the
scalar potential and non-linearities of the gravity, where the latter effect is
dominant than the former. The magnitude of the non-gaussianity in standard 
single-field inflation comes out to be small and of the order of slow-roll 
parameters \cite{Maldacena:2002vr}.

Inflation occurs at a very high energy  and it stretches out 
very small scales, of the order of Planck length, to the current hubble 
scale due to superluminal expansion of the universe. So. it provides
window to see the new physics at the Planck scale at which  
quantum corrections to the gravity becomes important. 
These new effects can significantly change the predictions of inflation  
that can be tested precisely by PLANCK experiment.    

Spacetime noncommutativity (see \cite{Akofor:2008ae} for review) is one of 
such modifications at high-energy, which
is well motivated by quantum gravity and string theory. Modifications to the power spectrum of scalar 
perturbations during inflation and its effects on CMB due to noncommutative
geometry has been studied in many places 
\cite{Lizzi:2002ib,Akofor:2007fv,Akofor:2008gv,Chu:2000ww,Brandenberger:2002nq}. 
In this paper we compute the three-point correlation functions of the 
curvature perturbation and hence the non-linearity parameter $f_{NL}$ 
determining the primordial non-gaussianity using the noncommutative quantum 
field theories related to deformed Poincare symmetry \cite{Akofor:2007fv}. 
In this approach of noncommutative geometry  quantum fields follow twisted 
statistics, as implied by the 
deformed Poincare symmetry in quantum theories. 
Non-gaussianity due to noncommutative geometry has been studied earlier in 
\cite{Fang:2007ba, Koivisto:2010fk}, where the former is based on the models motivated by 
string theory  and has considered the space-time components of noncommutativity parameter to be zero to
keep unitarily. while the latter has used the noncommutative spacetime with deformed Poincare 
symmetry as described in  \cite{Akofor:2007fv}. The computation of three-point function by 
Koivisto et al \cite{Koivisto:2010fk} is based on the $\delta N$ formalism 
\cite{Lyth:2005fi}, which is used to calculate the local non-gaussianity and treats the comoving
curvature perturbations as classical. But, here we compute the 
two-point and three-point function using Maldacena approach 
\cite{Maldacena:2002vr}  which is based on the second order perturbation theory
and takes the gravitational back reaction into account.

As described in \cite{Akofor:2007fv, Koivisto:2010fk}, the power spectrum of inflaton with 
noncommutative spacetime is direction dependent and can lead to the violation of statistical 
isotropy of CMB. PLANCK has seen some anomalies \cite{Ade:2013nlj}, specifically dipolar power 
modulation and hemispherical power asymmetry. Although the model studied in \cite{Akofor:2007fv} can 
not account for these anomalies, but generalization of it can lead to hemispherical power asymmetry
\cite{Groeneboom:2010fn}. 

The paper is organized as follows. In section \ref{qfnc}, after discussing spacetime noncommutativity 
we review the expressions for deformed quantum
fields and $\star$-product, described by Akofor et al \cite{Akofor:2007fv}, that are used to compute two and 
three-point correlation functions. 
In section \ref{ttp},  we review the  calculation of second and third-order action for comoving curvature
perturbation using ADM formalism and compute the power spectrum and three-point correlation function 
for the same in noncommutative Groenewold-Moyal plane. The expressions for the bispectrum and 
non-linearity parameter $f_{NL}$ with the three-point function obtained in section~\ref{ttp} are derived
in section~\ref{io} and there observational implications are also discussed. The conclusions are drawn
in section~\ref{cnc}.
  
\section{Quantum fields in noncommutative spacetime}
\label{qfnc}
At the energy scale of 
inflation, the noncommutativity of spacetime, which is motivated by
Heisenberg uncertainty principle and Einstein's general relativity, can play 
a crucial role.
The spacetime noncommutativity can be represented by the commutation 
relations \cite{Akofor:2008ae}
\be
\left[\tilde{x}_\mu,\tilde{x}_\nu\right]=i\theta_{\mu\nu} \label{noncom}
\ee
where $\theta_{\mu\nu}$ is a real antisymmetric matrix with constant elements 
and $\tilde{x}_\mu$ are the coordinate functions of the chosen coordinate system:
\be
\tilde{x}_\mu(x)=x_\mu.
\ee
The relation (\ref{noncom}) holds only in special coordinate systems and looks
quite complicated in other coordinates. The natural choice of the coordinate
for cosmological applications is the comoving frame, where the galaxies are
freely falling. This choice makes the time coordinate as the proper time measured by a clock 
at rest in any typical freely falling galaxy ($\vec{x}$ and $t$ are thus comoving coordinates) and also 
simplifies the calculations. 

Due to spacetime noncommutativity, the usual quantum fields are deformed and
can be given in terms of undeformed quantum fields as \cite{Akofor:2008ae}
\be
\phi_\theta=\phi_0 e^{\frac 12 \ola{\p}\wedge P}{\label{phitheta}}
\ee
where $\ola{\p}\wedge P=\ola{\p_\mu}\theta^{\mu\nu}P_\nu$ and 
$P_{\nu}$ represents the field momentum operator. The product of the 
deformed (twisted) quantum fields at the same spacetime point is 
represented by the star-product 
given as
\be
\left(\phi_\theta\star\phi_\theta\right)(x)=
\left.\phi_\theta(x) e^{\frac i2 \ola{\p_x}\wedge \ora{\p_y}}
\phi_\theta(y)\right|_{x=y}\, . \label{starp}
\ee
In the following sections we will make use of these relations to calculate
two and three-point correlation functions of the comoving curvature 
perturbations.
\section{Two-point and Three-point correlation functions with noncommutative spacetime}
\label{ttp}
\subsection{Background}
The action of a single scalar field minimally coupled with gravity is 
\be
S=\int d^4x \sqrt{-g}\left(\frac{{M_p}^2}{2}R+{\cal L}\right).\label{action}
\ee
Here $R$ is the Ricci scalar and ${\cal L}$ is the Lagrangian for the scalar 
field i.e.
\be
{\cal L}=-\frac{1}{2}g^{\mu\nu}\partial_\mu\phi\partial_\nu \phi-V(\phi).
\ee
Noncommutativity doesn't change the classical background so all the background
dynamics will be similar to the standard case. 
We take the metric signature $\left(-,+,+,+\right)$ and work in the units where
$M_p=1$. The background geometry of the homogeneous isotropic universe
is described by the  FRW metric
\be
ds^2=-dt^2+a^2(t)\left(dx^2+dy^2+dz^2\right).
\ee
For a scalar field dominated universe the  Friedmann equations are given as 
\bea
3H^2=\frac{1}{2}\dot\phi^2+V(\phi),\nonumber\\
\dot\rho+3 H(\rho+p)=0,\nonumber\\
\dot H=-\frac{1}{2}\left(\rho+p\right).\label{background}
\eea
The equation of motion for the scalar field is given as
\be
\ddot\phi+3H\dot\phi+\frac{dV(\phi)}{d\phi}=0.
\ee
During inflation, the potential energy of the scalar field dominates the total
energy density of the universe and the dynamics of the scalar field is 
 governed by slow-roll  parameters  defined as
as
\bea
\epsilon&=&-\frac{\dot H }{H^2},\nonumber \\
\eta &=& \frac{\dot\epsilon}{\epsilon H}. \label{slowrol}
\eea
Here we follow the definition of $\eta$ as in \cite{Seery:2005wm}, which is
different from the definition using scalar field potential 
($\eta_V=\frac{\frac{d^2V}{d\phi^2}}{V}$) and $\eta=-2\eta_V+4\ep$. 
\subsection{Perturbations and ADM formalism}
 The quantum fluctuations in the scalar field $\delta\phi(x,t)$
generated during inflation are coupled to the perturbations in the metric 
through Einstein's equation. Inflation gives rise to scalar and tensor 
perturbations in the metric and the scalar part is written as
\be
ds^2=-\left(1+2\Phi\right)dt^2+2a^2(t)B_{,i}dx^i dt +
a^2(t)\left(\left(1-2\Psi\right)\delta_{ij}+2E_{,ij}\right)dx^2dx^j.
\label{metricptb}
\ee
Here we have four scalar degrees of freedom in the metric and one in the 
scalar field which can be reduced to three by using gauge transformations.
We can again use the constraint equations derived from the 
perturbed Einstein's equation and describe the scalar perturbations in terms
of the curvature perturbation defined as \cite{Mukhanov:1990me}
\be
\zeta=-\Psi-\frac{H}{\dot\phi}\delta\phi.\label{zetadefinition}
\ee
This variable is gauge invariant and is conserved on super-horizon scales.
One can write the action (\ref{action}) in terms of $\zeta$ and it turns out
quadratic in $\zeta$.
To do the perturbation theory in higher order it is convenient to  use ADM 
formalism where the metric can be written as \cite{Arnowitt:1962hi}
\be
ds^2=-N^2dt^2+h_{ij}\left(dx^i+N^i dt\right)\left(dx^j+N^j dt\right)
\label{admmetric}
\ee
where $N$ is laps function,  $N_i,\, N_j$ are shift vectors and $h_{ij}$ is
metric of three-dimensional hypersurface of constant time. Here $N$ and $N_i$
appear as Lagrangian multipliers in the action so one can solve there 
constraint equations  and substitute the solution back into the action.
This simplifies the tedious calculations needed while working with 
(\ref{metricptb}).
Now we  chose comoving gauge $\delta \phi=0$ to do our calculation and
in this gauge we can use non-linear generalization of $\zeta$ 
\cite{Lyth:2005du} and 
define the gauge as \cite{Maldacena:2002vr,Seery:2005wm}
\be
h_{ij}=a^2 e^{2\zeta}\delta_{ij},\, \, \delta\phi=0. \label{gauge}
\ee
With this gauge choice the action (\ref{action}) with the 
metric (\ref{admmetric}) becomes
\be
S=\frac{1}{2}\int dt d^3x\sqrt{h}\left(N R^{(3)}-2N V(\phi)+
N^{-1}\dot\phi^2+N^{-1}\left(E_{ij}E^{ij}-E^2\right)\right).\label{admaction}
\ee
Here $R^{(3)}$ represents the Ricci scalar calculated using the 
three-dimensional metric $h_{ij}$ and $E_{ij}$ is related to the extrinsic
curvature of the constant time hypersurface and is given as  
\be
E_{ij}=\frac{1}{2}\left(\dot h_{ij}-\nabla_j N_i-\nabla_i N_j\right).
\ee
Varying the action (\ref{admaction})  we get the constraint equation  for $N$ 
and $N^i$ as 
\bea
R^{(3)}-2 V-N^{-2}(E_{ij}E^{ij}-E^2)-N^{-2}\dot\phi^2=0,\nonumber\\
\nabla_j\left[N^{-1}\left(E_i^j-\delta_i^jE\right)\right]=0.\label{constraint}
\eea 
Now we can decompose $N_i$ into irrotational and incompressible parts as 
$N_i=\tilde{N_i}+\partial_i\psi$ where $\partial_i\tilde{N^i}=0$ and expand 
$N$, $\psi$ and $\tilde{N^i}$ into powers of $\zeta$ as
\bea
N&=&1+\alpha_1+\alpha_2+.....\, ,\nonumber\\
\tilde{N_i}&=&\tilde{N_i}^{(1)}+\tilde{N_i}^{(2)}+...\, ,\nonumber\\
\psi&=&\psi_1+\psi_2+......\, . \label{expand}
\eea
Using these expansions, the constraint equations (\ref{constraint}) can be
solved order by order with metric (\ref{gauge}) and at first order one gets
\be
\alpha_1=\frac{\dot\zeta}{H},\, \tilde{N_i}^{(1)}=0,\, 
\psi_1=-\frac{\zeta}{H}+\chi,\, \partial^2\chi=a^2\epsilon \dot\zeta. 
\label{sol}
\ee
Here $\partial^2=\delta^{ij}\partial_i\partial_j$ and the use of suitable 
choice of boundary conditions has been made to put $N_i^{(1)}=0$. 
As mentioned in \cite{Maldacena:2002vr, Seery:2005wm} to calculate the 
action up to $n^{\rm {th}}$ order in $\zeta$, we need to calculate $N$ and $N_i$ only
up to the order-$\zeta^{n-1}$ and here the terms of order-$\zeta^2$ also drop
out from the third order action, so equation (\ref{sol}) is sufficient to 
compute the action up to third order.
So, after putting these solutions in (\ref{admaction}) we get the action for 
second and third order in $\zeta$ as 
\cite{Maldacena:2002vr,Chen:2006nt,Seery:2005wm}
\bea
S_2&=&\int dtd^3x \left[a^3\epsilon \dot\zeta^2-
a\epsilon(\partial \zeta)^2\right],\label{s2}\\
S_3&=&\int dtd^3x\left[-a\epsilon\zeta(\partial\zeta)^2-
a^3\epsilon\dot\zeta^3+3a^3\epsilon\zeta\dot\zeta^2\right.\nonumber\\
 & &\left. +\frac{1}{2 a}\left(3\zeta-\frac{\dot\zeta}{H}\right)
\left(\partial_i\partial_j\psi\partial^i\partial^j\psi-
\partial^2\psi\partial^2\psi\right)-2a^{-1}\partial_i\psi\partial_i\zeta
\partial^2\psi\right].\label{s3}
\eea
\subsection{Two-point correlation function and power spectrum}
Now to calculate the two-point correlation function the quadratic part 
(\ref{s2}) of the action is considered, which in conformal time
($d\tau=\frac{dt}{a}$) can be written as 
\be
S_2=\int d\tau d^3x a^2\epsilon \left[\zeta^{\prime 2}-
(\partial \zeta)^2\right].\label{sconformal}
\ee
Here $\pr$ denotes derivative w.r.t conformal time $\tau$. The above action
 looks like an action of a massless scalar field in conformal spacetime
and  $\zeta$ can be considered as the scalar field for quantization.
 $\zeta$ can be written in terms of creation and annihilation
operator as
\be
\zeta(\vec{x},\tau)=\int \frac{d^3k}{(2\pi)^3}\zeta(\vec{k},\tau)
e^{i\vec{k}\cdot\vec{x}}=\int \frac{d^3k}{(2\pi)^3}
\left(u(\vec{k},\tau)a_{\vec{k}}+u^\star(-\vec{k},\tau)a_{-\vec{k}}^\dagger\right)
e^{i\vec{k}\cdot\vec{x}}.\label{zetafourier}
\ee
The equation of motion for $\zeta$ can be obtained by varying the action 
(\ref{sconformal}) and is given by
\be
\zeta^{\prime\prime}+2\frac{z^\prime}{z}\zeta^\prime-\partial^2\zeta=0.
\label{zetaeqm}
\ee
Here $z^2=2 a^2\epsilon$ and we can define 
$v_{\vec{k}}=z\zeta(\vec{k},\tau)$ and use equation (\ref{zetafourier}) to 
get
\be
v_{\vec{k}}^{\prime\prime}+\left(k^2-\frac{z^{\prime\prime}}{z}\right)v_{\vec{k}}=0.
\label{veqm}
\ee
The solution for the mode functions $v_{\vec{k}}$ can be obtained assuming 
Bunch Davies initial conditions and is given as
\be
v_{\vec{k}}=\frac{1}{\sqrt{2k}}\left(1-\frac{i}{k\tau}\right)e^{-i k\tau}.\label{vsol}
\ee
Hence the basis function $u(\vec{k},\tau)$ is 
\be
u(\vec{k},\tau)=\frac{v_{\vec{k}}}{z}=\frac{i H}{\sqrt{4\epsilon k^3}}
\left(1+ik\tau\right)e^{-ik\tau}.\label{uksol}
\ee
The two point correlation function of the field $\zeta$ in position space
can be expressed  as
\bea
\langle\zeta(\vec{x},\tau)\zeta(\vec{y},\tau)\rangle&=&
\int \frac{d^3k d^3k^\prime}{(2\pi)^6}
\langle 0|\zeta(\vec{k},\tau)\zeta(\vec{k^\prime},\tau)|0\rangle 
e^{i\left(\vec{k}\cdot\vec{x}+\vec{k^\prime}\cdot\vec{y}\right)}
\label{twopoint}\\
&=&\int \frac{d^3k}{(2\pi)^3}{|u(\vec{k},\tau)|}^2
e^{i\vec{k}\cdot\left(\vec{x}-\vec{y}\right)},
\eea
where we have used the relation 
$\langle 0|\zeta(\vec{k},\tau)\zeta(\vec{k^\prime},\tau)|0\rangle=
(2\pi)^3 \delta^3\left(\vec{k}+\vec{k}^\pr\right)u(\vec{k},\tau)
u^\star(-\vec{k}^\pr,\tau)$. The power spectrum for $\zeta$ is 
defined by 
\be
\lan 0|\zeta(\vec{k},\tau)\zeta(\vec{k^\pr},\tau)|0\ran=(2\pi)^3
\delta^3\left(\vec{k}+\vec{k}^\pr\right)P_{\zeta}(k).\label{powerd}
\ee
So
\be
P_{\zeta}(k)={|u(\vec{k},\tau)|}^2.\label{power}
\ee
The another convention for the power spectrum, that is commonly used for 
data analysis, is
\be
\Delta_{\zeta}^2=\frac{k^3}{2\pi^2}{|u(\vec{k},\tau)|}^2.\label{variance}
\ee 
In this case $\Delta_\zeta$ represents the variance of the classical 
fluctuations and the two-point correlation in position space becomes
\be
\langle\zeta(\vec{x},\tau)\zeta(\vec{y},\tau)\rangle=
\int \frac{dk}{k}\Delta_{\zeta}^2e^{i\vec{k}\cdot\left(\vec{x}-\vec{y}\right)}.
\ee
The power spectrum is calculated on super-horizon limit i.e. $-k\tau<<1$ 
in which $v_{\vec{k}}=\frac{1}{\sqrt{2k}}\left(-\frac{i}{k\tau}\right)e^{-i k\tau}$ and we get the power spectrum  as 
\be
P_{\zeta}(k)=\frac{H^2}{4\epsilon}\frac{1}{k^3}.\label{powers}
\ee
Now due to noncommutativity of spacetime the two point correlation
function for field $\zeta$ gets modified \cite{Akofor:2007fv}. We will denote 
the  field in noncommutative spacetime with a subscript $\theta$. 
Since here $\zeta$ represents our quantum field, hence similar to 
(\ref{phitheta}) the twisted quantum field $\zeta_\theta$ can be expressed in 
terms of the untwisted field $\zeta$  as
\be
\zeta_{\theta}(\vec{x},t)= \zeta(\vec{x},t)
e^{\frac12\overleftarrow\partial_\mu\wedge P_\nu}.\label{zetatheta}
\ee
With the twisted quantum field one can compute the two-point correlation
function in position space as 
\bea
\langle\zeta_\theta(\vec{x},t)\zeta_\theta(\vec{y},t^\prime)\rangle&=&
\langle\zeta(\vec{x},t)e^{\frac12\overleftarrow\partial_{x_\mu}\wedge P_\nu}
\zeta(\vec{y},t^\prime)e^{\frac12\overleftarrow\partial_{y_\mu}\wedge P_\nu}
\rangle\nonumber\\
&=&\langle\zeta(\vec{x},t)\zeta(\vec{y},t^\prime)\rangle
e^{-\frac{i}{2}\overleftarrow 
\partial_{x_\mu}\wedge\overrightarrow\partial_{y_\nu}},
\eea
where we have used the commutation relations between the field and the 
momentum operator $\left[P_\mu,\zeta\right]=-i\p_\mu\zeta$. Now taking the 
Fourier transform on the right hand side we get
\bea
\langle\zeta_\theta(\vec{x},t)\zeta_\theta(\vec{y},t^\prime)\rangle
&=&\int \frac{d^3k d^3k^\prime}{(2\pi)^6}
\langle 0|\zeta(\vec{k},t)\zeta(\vec{k^\prime},t^\prime)|0\rangle
e^{-\frac{i}{2}\overleftarrow \partial_{x_\mu}\wedge\overrightarrow\partial_{y_\nu}}
e^{i\left(\vec{k}\cdot\vec{x}+\vec{k^\prime}\cdot\vec{y}\right)}\nonumber\\
&=&\int \frac{d^3k d^3k^\prime}{(2\pi)^6}
\langle 0|\zeta(\vec{k},t)\zeta(\vec{k^\prime},t^\prime)|0\rangle
e^{-\frac{i}{2}\left(\partial_t\theta^{0i}\partial_{\vec{y}}+
\partial_{\vec{x}}\theta^{io}\partial_{t^\prime}+\partial_{\vec{x}}\wedge
\partial_{\vec{y}}\right)}e^{i\left(\vec{k}\cdot\vec{x}+\vec{k^\prime}\cdot\vec{y}\right)}\nonumber\\
&=&\int \frac{d^3k d^3k^\prime}{(2\pi)^6}
\langle 0|\zeta(\vec{k},t)\zeta(\vec{k^\prime},t^\prime)|0\rangle
e^{\left(\frac{i}{2}\vec{k}\wedge\vec{k^\prime}+
\frac{\ora{\theta^0}\cdot\vec{k^\prime}}{2}\partial_t-
\frac{\ora{\theta^0}\cdot\vec{k}}{2}\partial_{t^\prime}\right)}
e^{i\left(\vec{k}\cdot\vec{x}+\vec{k^\prime}\cdot\vec{y}\right)}\nonumber\\
&=&\int \frac{d^3k d^3k^\prime}{(2\pi)^6}
\langle 0|\zeta\left(\vec{k},t+\frac{\ora{\theta^0}\cdot\vec{k^\prime}}{2}
\right)\zeta\left(\vec{k^\prime},t^\prime-\frac{\ora{\theta^0}\cdot\vec{k}}{2}
\right)|0 \rangle 
e^{\frac{i}{2}\vec{k}\wedge\vec{k^\prime}}
e^{i\left(\vec{k}\cdot\vec{x}+\vec{k^\prime}\cdot\vec{y}\right)}.\nonumber\\
\label{twopointtheta}
\eea
Here  $\ora{\theta^0}=\theta^{0i}$. So the two-point correlation function
in momentum space can be expressed as
\be
\lan0|\zeta_\theta(\vec{k},t)\zeta_\theta(\vec{k^\pr},t^\pr)|0\ran=
e^{\frac{i}{2}\vec{k}\wedge\vec{k^\prime}}
\langle 0|\zeta\left(\vec{k},t+\frac{\ora{\theta^0}\cdot\vec{k^\prime}}{2}
\right)\zeta\left(\vec{k^\prime},t^\prime-\frac{\ora{\theta^0}\cdot\vec{k}}{2}
\right)|0 \rangle.
\ee
Now since in de Sitter space
\be
\tau(t)=\frac{1}{a H} e^{-H t}. \label{taut}
\ee
So in conformal time and in the limit $t^\prime \rightarrow t$
\bea
\zeta\left(\vec{k},t+\frac{\ora{\theta^0}\cdot\vec{k^\prime}}{2}\right)
&\rightarrow&
\zeta\left(\vec{k},\tau e^{-H \frac{\ora{\theta^0}\cdot\vec{k^\prime}}{2}}\right)\, ,\\
\zeta\left(\vec{k},t^\prime-\frac{\ora{\theta^0}\cdot\vec{k}}{2}\right)
&\rightarrow&
\zeta\left(\vec{k},\tau e^{H \frac{\ora{\theta^0}\cdot\vec{k}}{2}}\right).
\eea
Hence the two-point function will be
\bea
\langle\zeta_\theta(\vec{k},\tau)\zeta_\theta(\vec{k^\pr},\tau)\ran&=&
\lan 0|\zeta\left(\vec{k},\tau e^{-H \frac{\ora{\theta^0}\cdot\vec{k^\prime}}
{2}}\right)
\zeta\left(\vec{k^\pr},\tau e^{H \frac{\ora{\theta^0}\cdot\vec{k}}{2}}\right)
|0 \rangle
e^{\frac{i}{2}\vec{k}\wedge\vec{k^\prime}}\nonumber\\
&=&{\left|u\left(\vec{k},\tau e^{H \frac{\ora{\theta^0}\cdot\vec{k}}{2}}\right)\right|}^2(2\pi)^3\delta^3(\vec{k}+\vec{k^\pr}).
\eea
Now we take the self-adjoint part of two-point correlation function defined as
\cite{Akofor:2007fv}
\be
{\lan0|\zeta_\theta(\vec{k},\tau)\zeta_\theta(\vec{k^\pr},\tau)|0\ran}_M=
\frac12\left(\lan0|\zeta_\theta(\vec{k},\tau)\zeta_\theta(\vec{k^\pr},\tau)
|0\ran+\lan0|\zeta_\theta(-\vec{k},\tau)\zeta_\theta(-\vec{k^\pr},\tau)\right).
\ee
So the power spectrum can be obtained from (\ref{powerd}) as
\be
P_{\zeta_\theta}(k)=\frac12\left(
{\left|u\left(\vec{k},\tau e^{H \frac{\ora{\theta^0}\cdot\vec{k}}{2}}\right)\right|}^2+{\left|u\left(-\vec{k},\tau e^{-H \frac{\ora{\theta^0}
\cdot\vec{k}}{2}}\right)\right|}^2\right).
\ee
Now since $v_{\vec{k}}=z \zeta(\vec{k},\tau)$, the argument of $v_{\vec{k}}$ is shifted due to 
deformation of $\zeta(\vec{k},\tau)$ and the argument of the scale factor $a(\tau)$ and hence $z$ is not
shifted.

Since on super-horizon limit $v_{\vec{k}}=\frac{1}{\sqrt{2 k}}
\left(\frac{-i}{k\tau}e^{-H \frac{\ora{\theta^0}\cdot\vec{k}}{2}}\right) $ so 
\be
P_{\zeta_\theta}(k)=P_{\zeta}(k)\cosh\left({H \ora{\theta^0}\cdot\vec{k}}
\right).\label{powernc}
\ee
This power spectrum was derived in \cite{Akofor:2007fv,Koivisto:2010fk} and they showed that 
it can lead to the breaking of statistical isotropy of the CMB. Akofor et al. \cite{Akofor:2008gv} tested
the above power spectrum with WMAP5 \cite{Nolta:2008ih}, ACBAR \cite{Reichardt:2008ay} and 
CBI \cite{Mason:2002tm} data sets considering only the effects on $Cl$s and
ignoring the off-diagonal terms in $\langle a_{lm} a_{l^\prime m^\prime}\rangle$ correlations. As
the effects of modifications to the power spectrum due to noncommutativity increase at small scales,
it was concluded in \cite{Akofor:2008gv}  that WMAP5 data, which gives the power spectra for $Cl$s
up to $l=1000$,  is not sufficient to constrain the scale of noncommutativity. Doing a 
one-parameter $\chi^2$ analysis with ACBAR   and CBI data, which give CMB power spectra up to $l=2958$ 
and $l=3500$ respectively (but only for small scales), they claimed that $H\theta^0<0.01$MPc 
(where $\theta^0$ is the magnitude of the noncommutativity parameter $\ora{\theta^0}$). Recently PLANCK has released data for the CMB power spectra up to $l=2500$ \cite{Planck:2013kta}
with better precision and less systematic errors. 
Since there may be parameter degeneracy (for e.g due to spectral index), we are 
planning to reanalyze the power spectrum (\ref{powernc}) with the recently released PLANCK data by 
varying all parameters along with $H\theta^0$ to constraint the scale of noncommutativity.

The above power spectrum can be
expanded in terms of $\left({H \ora{\theta^0}\cdot\vec{k}}\right)$ and keeping only the leading order 
term we get,
\be
 P_{\zeta_\theta}(k)=P_{\zeta}(k)\left(1+\frac{\left(H\theta^0 k\right)^2}{2}
\left({\hat{\theta^0}\cdot\hat{k}}\right)^2\right),
\label{powerex}
\ee
here $k$ denote the magnitudes of the  wavenumber 
and $\hat{\theta^0}$ is a unit vector in the direction of $\ora{\theta^0}$ along  which 
the rotational invariance is broken. 
A power spectrum of similar form was considered in \cite{Ackerman:2007nb} where a small non-zero 
vector was introduced to break the rotational invariance and the coefficient of the  
direction dependent term (denoted by $g(k)$ in \cite{Ackerman:2007nb}) was scale invariant. 
Groeneboom et al. \cite{Groeneboom:2008fz} analyzed the power spectrum of \cite{Ackerman:2007nb} with 
WMAP5 data and obtained the bound $g=0.29\pm0.031$ with the exclusion of $g=0$ at $9\sigma$ by 
including the CMB multipoles up to $l=400$. The result was contradicted by 
Hanson et al. \cite{Hanson:2010gu} and they argued 
that  the detection of non-zero $g$ can be due to the beam asymmetry. Pullen et al. 
\cite{Pullen:2010zy} re-analyzed the power spectrum of \cite{Ackerman:2007nb} with the 
large scale structure surveys and they obtained $g=0.007 \pm 0.037$.
The power spectra (\ref{powernc}) and  of Ackerman et al. \cite{Ackerman:2007nb} give rise to multipole
alignments along the preferred direction. The quadrupole-octopole alignment was first reported by
Tegmark et al. \cite{Tegmark:2003ve} using the WMAP first year data and was less significant in WMAP.
With recently released PLANCK data, the significance for multipole alignment is even smaller than WMAP
\cite{Ade:2013nlj}. The off-diagonal terms in  $\langle a_{lm} a_{l^\prime m^\prime}\rangle$ arising due
to the power spectrum (\ref{powernc}) can be described by bipolar spherical harmonics 
(BipoSH)\cite{Hajian:2003qq} representing the modulation of the CMB power spectrum. 
PLANCK claims 3.7 to 2.9$\sigma$ detection of dipole modulation (non-zero $L=1$ BipoSH)  
but null result for higher multipoles of BipoSH. The power spectrum (\ref{powernc}) can only give rise to
even multipole BiopoSH so it can not account for the observed dipole modulation of CMB.    
We will describe the modified three-point correlation function due to noncommutativity and its 
observational implications in the next sections.
 
\subsection{Three-point function}
The primordial non-gaussianity in CMB arises due to the non-zero three-point
and four-point correlation functions of curvature perturbations. 
These correlation functions were calculated for noncommutative spacetime in 
\cite{Koivisto:2010fk}, where they have used $\delta N$ formalism which 
ignores modifications to the correlation functions at Hubble crossing and also
interaction between quantum fluctuations on sub-hubble scales with the 
super-hubble scale fluctuations at non-linear label.  

The third order action  (\ref{s3})  obtained for $\zeta$ using ADM formalism 
is
\bea
S_3&=&\int dtd^3x\left[-a\epsilon\zeta(\partial\zeta)^2-
a^3\epsilon\dot\zeta^3+3a^3\epsilon\zeta\dot\zeta^2\right.\nonumber\\
 & &\left. +\frac{1}{2 a}\left(3\zeta-\frac{\dot\zeta}{H}\right)
\left(\partial_i\partial_j\psi\partial^i\partial^j\psi-
\partial^2\psi\partial^2\psi\right)-2a^{-1}\partial_i\psi\partial_i\zeta
\partial^2\psi\right].
\eea
We  put the value of $\psi$ from Eq.~(\ref{sol}) in this action, 
 integrate by parts  and use background Friedmann equations to get terms 
proportional to $\ep^2$
\bea
S_3&=&\int dt d^3x \left[a^3\ep^2\zeta\dot\zeta^2+a\ep^2\zeta(\p\zeta)^2
-2a\ep\dot\zeta(\p\zeta)(\p\chi)+
\frac{a^3\ep}{2}\frac{d\eta}{dt}\zeta^2\dot\zeta\right.\nonumber\\
&+&\left. \frac{\ep}{2a}(\p\zeta)(\p\chi)(\p^2\chi)
+\frac{\ep}{4a}(\p^2\zeta)(\p\chi)^2+
\frac{1}{2}a {\cal F}{\left.\frac{\delta L}{\delta\zeta}\right|}_1\right]
\eea
where ${\cal F}= \left(\eta\zeta^2+\, \text{terms with derivatives of}\, \zeta\right)$ and $\frac{\delta L}{\delta\zeta}$ represents the terms proportional to the Gaussian action $S_2$.
We can again integrate by parts the above action to remove the terms involving
$\p\chi$ and use the Gaussian field equation (\ref{zetaeqm}) to get
\be
S_3=\int dt d^3x\left[4 a^5 \epsilon^2 H \dot\zeta^2\partial^{-2}\dot\zeta
+\frac{1}{2}a {\cal F}{\left.\frac{\delta L}{\delta\zeta}\right|}_1 \right].
\label{s3one}
\ee
Now ${\cal F}=\left(\eta-\epsilon\right)\zeta^2+
2\epsilon \partial^{-2}\left(\zeta\partial^2\zeta\right)$ and 
$\partial^{-2}$ is the inverse of $\partial^2$ and we have ignored the 
terms containing the derivatives of $\zeta$ in ${\cal F}$ as they are 
negligible on super-horizon scales. One can get rid of the second term in the above
action following field redefinition
\be
\zeta\rightarrow \zeta_n+\frac{{\cal F}}{4}\zeta_n^2. \label{zetaredefine}
\ee
After this field redefinition the three-point function becomes 
\bea
\langle\zeta(x_1)\zeta(x_2)\zeta(x_3)\rangle&=& 
\langle\zeta_n(x_1)\zeta_n(x_2)\zeta_n(x_3)\rangle
+\frac{\left(\eta-\epsilon\right)}{4}\left(\langle \zeta_n(x_1)
\zeta_n(x_2)\rangle
\langle\zeta_n(x_1)\zeta_n(x_3)\rangle+\, \text{permutations}\right)\nonumber\\ &+&
\frac{\epsilon}{2}\partial^{-2}_{x_1}\left(\langle\zeta(x_1)\zeta(x_2)\rangle
\partial^2_{x_1}\langle\zeta_n(x_1)\zeta_n(x_3)\rangle+\, \text{permutations}\right).
\label{zetathreepoint}
\eea
The first term in above expression represents the three-point function while the last
two terms represents the corrections to the three-point function due to field 
redefinition. We will omit the subscript $n$ in the following calculations. 
Now  the interaction Hamiltonian to calculate the three-point function can be 
obtained from action (\ref{s3one}) i.e
\be
{\cal H}(t^\prime)=-\int d^3 x 4 a^5 \epsilon^2 H \dot\zeta^2\partial^{-2}\dot\zeta.
\label{ht}
\ee
As mentioned earlier, we use $\zeta$ as the quantum field to compute the 
various correlation functions. Hence, to see the effects of noncommutative geometry on three-point 
correlation function we replace the usual quantum field $\zeta$ with the twisted field
$\zeta_\theta$ both in the interaction Hamiltonian and in (\ref{zetathreepoint}). 
Since the product of the twisted fields at the same spacetime point is given 
by the star-product \cite{Akofor:2008ae}, the interaction Hamiltonian will be given 
as
\be
{\cal H}(t^\prime)=-\int d^3 x 4 a^5 \epsilon^2 H 
\dot\zeta_\theta\star \dot\zeta_\theta\star\partial^{-2}\dot\zeta_\theta.
\ee
One important point to be mentioned here is that, in principle, we should replace 
$\zeta$ with $\zeta_\theta$ and the product between them as star-product in equation (\ref{s3}), but 
since $\theta_{\mu\nu}$ is constant in comoving coordinates and hence the star-product of the deformed
fields is associative \cite{Akofor:2008ae}, all the steps to reach to interaction Hamiltonian from
the third-order action can be performed as in standard case and $\zeta$ can be replaced with 
$\zeta_\theta$ in the final interaction Hamiltonian.  

Using the relation (\ref{phitheta}) between the twisted and untwisted quantum
field and expression for star-product (\ref{starp}), the interaction 
Hamiltonian becomes 
\be
{\cal H}(t^\prime)=-\int d^3 x 4 a^5 \epsilon^2 H
\dot\zeta^2\partial^{-2}\dot\zeta e^{\frac12\ola{\p_{\mu}}\wedge P_\nu},
\label{hinttheta}
\ee
where $\ola{\p_{\mu}}\wedge P_\nu=\ola{\p_{x_\mu}}\theta^{\mu\nu} P_\nu$. The first term in the RHS of (\ref{zetathreepoint}) is computed using the 
in-in formalism \cite{Weinberg:2005vy} and is given by
\bea
\langle\zeta_\theta(x_1)\zeta_\theta(x_2)\zeta_\theta(x_3)\rangle&=&
-i\int_{t_0}^t dt^\pr\lan 0|\left[\zeta_\theta(x_1)\zeta_\theta(x_2)\zeta_\theta(x_3),
{\cal H}(t^\pr)\right]|0\ran\nonumber\\
&=&-i \int_{t_0}^t dt^\pr \left(\lan 0|\zeta_\theta(x_1)\zeta_\theta(x_2)
\zeta_\theta(x_3){\cal H}(t^\pr)|0\ran-\lan 0|{\cal H}(t^\pr)\zeta_\theta(x_1)\zeta_\theta(x_2)\zeta_\theta(x_3)|0\ran\right).\nonumber\\\label{zetathree1}
\eea 
Here the three-point function is calculated at equal time i.e $t_1=t_2=t_3=t$. Initially we will
write them differently for simplification but will put them equal before integration w.r.t 
$t^\pr$.
Let us now consider the first term of above equation with (\ref{hinttheta}) 
and name it as $(a)$. So
\be
(a)=4 i\ep^2\int dt^\pr a^5 H \int d^3x
\lan 0|\zeta_\theta(x_1)\zeta_\theta(x_2)\zeta_\theta(x_3)
{\left.\dot\zeta^2\p^{-2}\dot\zeta\right|}_{t^\pr,\vec{x}}
e^{\frac12\ola{\p_{x_\mu}}\wedge P_\nu}|0\ran.
\ee
Using the relation (\ref{zetatheta}) we can replace the twisted quantum fields
in terms of the untwisted quantum fields and it gives
\bea
(a)&=&4 i\ep^2\int dt^\pr a^5 H \int d^3x
\lan 0|\zeta(x_1)\zeta(x_2)\zeta(x_3)
e^{-\frac i2\left(\ola{\p_{x_1}}\wedge \ola{\p_{x_2}}+
\ola{\p_{x_2}}\wedge \ola{\p_{x_3}}+\ola{\p_{x_1}}\wedge \ola{\p_{x_3}}\right)}
\nonumber\\
&\times&e^{\frac12\ola{\p_{x_1}}\wedge P}e^{\frac12\ola{\p_{x_2}}\wedge P}
e^{\frac12\ola{\p_{x_3}}\wedge P}
{\left.\dot\zeta^2\p^{-2}\dot\zeta\right|}_{t^\pr,\vec{x}}
e^{\frac12\ola{\p_{x}}\wedge P} |0\ran\\
&=&4 i\ep^2\int dt^\pr a^5 H \int d^3x
\lan 0|\zeta(x_1)\zeta(x_2)\zeta(x_3)
e^{-\frac i2\left(\ola{\p_{x_1}}\wedge \ola{\p_{x_2}}+
\ola{\p_{x_2}}\wedge \ola{\p_{x_3}}+\ola{\p_{x_1}}\wedge \ola{\p_{x_3}}\right)}
\nonumber\\
&\times&e^{-\frac i2\left(\ola{\p_{x_1}}+\ola{\p_{x_2}}+\ola{\p_{x_3}}\right)
\wedge \ora{\p_x}} \dot\zeta 
e^{-\frac i2\left(\ola{\p_{x_1}}+\ola{\p_{x_2}}+\ola{\p_{x_3}}\right)
\wedge \ora{\p_x}} \dot\zeta
e^{-\frac i2\left(\ola{\p_{x_1}}+\ola{\p_{x_2}}+\ola{\p_{x_3}}\right)
\wedge \ora{\p_x}}\p^{-2} \dot\zeta|0\ran.\nonumber\\
\eea
The above equation in Fourier space becomes
\bea
(a)&=&-4i\ep^2\int dt^\pr a^5 H \int d^3x\int\prod_{i=1}^6\frac{d^3k_i}{k^2_6
(2\pi)^{18}}
e^{i\left(\vec{k}_1\cdot\vec{x}_1+\vec{k}_2\cdot\vec{x}_2+\vec{k}_3\cdot\vec{x}_3\right)}
\nonumber\\
&\times&\lan 0|\zeta\left(\vec{k}_1,t_1+\frac{\ora{\theta^0}\cdot\vec{k}_2+
\ora{\theta^0}\cdot\vec{k}_3+\ora{\theta^0}\cdot\vec{k}_4+\ora{\theta^0}\cdot\vec{k}_5+
\ora{\theta^0}\cdot\vec{k}_6}{2}\right)\nonumber\\
&\times&\zeta\left(\vec{k}_2,t_2+\frac{-\ora{\theta^0}\cdot\vec{k}_1+
\ora{\theta^0}\cdot\vec{k}_3+\ora{\theta^0}\cdot\vec{k}_4+\ora{\theta^0}\cdot\vec{k}_5+
\ora{\theta^0}\cdot\vec{k}_6}{2}\right)\nonumber\\
&\times&\zeta\left(\vec{k}_3,t_3+\frac{-\ora{\theta^0}\cdot\vec{k}_1-
\ora{\theta^0}\cdot\vec{k}_2+\ora{\theta^0}\cdot\vec{k}_4+\ora{\theta^0}\cdot\vec{k}_5+
\ora{\theta^0}\cdot\vec{k}_6}{2}\right)\nonumber\\
&\times&\dot\zeta\left(\vec{k}_4,t^\pr-\frac{\ora{\theta^0}\cdot\vec{k}_1+
\ora{\theta^0}\cdot\vec{k}_2+\ora{\theta^0}\cdot\vec{k}_3}{2}\right)\nonumber\\
&\times&\dot\zeta\left(\vec{k}_5,t^\pr-\frac{\ora{\theta^0}\cdot\vec{k}_1+
\ora{\theta^0}\cdot\vec{k}_2+\ora{\theta^0}\cdot\vec{k}_3}{2}\right)\nonumber\\
&\times&\dot\zeta\left(\vec{k}_6,t^\pr-\frac{\ora{\theta^0}\cdot\vec{k}_1+
\ora{\theta^0}\cdot\vec{k}_2+\ora{\theta^0}\cdot\vec{k}_3}{2}\right)
|0\ran\nonumber\\
&\times& e^{i\left(\vec{k}_4\cdot\vec{x}+\vec{k}_5\cdot\vec{x}+\vec{k}_6\cdot\vec{x}\right)}
e^{\frac i2 {\cal P}}.\label{a1}
\eea
Here 
\be
{\cal P}=\left(\vec{k}_1\wedge\vec{k}_2+\vec{k}_2\wedge\vec{k}_3+\vec{k}_1\wedge\vec{k}_3+
\left(\vec{k}_1+\vec{k}_2+\vec{k}_3\right)\left(\vec{k}_4+\vec{k}_5+\vec{k}_6\right)\right).
\label{phase}
\ee
Since we will express the three-point correlation function in momentum space,
we can take the Fourier transform on both side of Eq.~(\ref{zetathree1}) and 
take the limit $t_1=t_2=t_3=t$ to get 
\bea
(a)&=&-i\int_{t_0}^t dt^\pr\left(\lan 0|\zeta_\theta(\vec{k}_1,t)
\zeta_\theta(\vec{k}_2,t)\zeta_\theta(\vec{k}_3,t){\cal H}(t^\pr)\right)
\nonumber\\
&=&-4i\ep^2\int dt^\pr a^5 H \int d^3x\int\prod_{i=4}^6\frac{d^3k_i}{k^2_6(2\pi)^9}
\lan 0|\zeta(\vec{k}_1,t_1)\zeta(\vec{k}_2,t_2)\zeta(\vec{k}_3,t_3)
\dot\zeta(\vec{k}_4,t_4)\dot\zeta(\vec{k}_5,t_5)\dot\zeta(\vec{k}_6,t_6)|0\ran\nonumber\\
&\times&e^{i\left(\vec{k}_4\cdot\vec{x}+\vec{k}_5\cdot\vec{x}+\vec{k}_6\cdot\vec{x}\right)}
e^{\frac i2 {\cal P}}\label{a2}
\eea
where
\bea
t_1&=&t+\frac{\ora{\theta^0}\cdot\vec{k}_2+
\ora{\theta^0}\cdot\vec{k}_3+\ora{\theta^0}\cdot\vec{k}_4+\ora{\theta^0}\cdot\vec{k}_5+
\ora{\theta^0}\cdot\vec{k}_6}{2},\nonumber\\
t_2&=&t+\frac{\ora{-\theta^0}\cdot\vec{k}_1+
\ora{\theta^0}\cdot\vec{k}_3+\ora{\theta^0}\cdot\vec{k}_4+\ora{\theta^0}\cdot\vec{k}_5+
\ora{\theta^0}\cdot\vec{k}_6}{2},\nonumber\\
t_3&=&t+\frac{\ora{-\theta^0}\cdot\vec{k}_1-
\ora{\theta^0}\cdot\vec{k}_2+\ora{\theta^0}\cdot\vec{k}_4+\ora{\theta^0}\cdot\vec{k}_5+
\ora{\theta^0}\cdot\vec{k}_6}{2},\nonumber\\
t_4&=&t^\pr-\frac{\ora{\theta^0}\cdot\vec{k}_1+
\ora{\theta^0}\cdot\vec{k}_2+\ora{\theta^0}\cdot\vec{k}_3}{2},\nonumber\\
t_5&=&t^\pr-\frac{\ora{\theta^0}\cdot\vec{k}_1+
\ora{\theta^0}\cdot\vec{k}_2+\ora{\theta^0}\cdot\vec{k}_3}{2},\nonumber\\
t_6&=&t^\pr-\frac{\ora{\theta^0}\cdot\vec{k}_1+
\ora{\theta^0}\cdot\vec{k}_2+\ora{\theta^0}\cdot\vec{k}_3}{2}.\label{time}\\
\eea
A detailed calculation of this term is presented in Appendix and it is 
given as (\ref{aapdsol})
\be
(a)=\ep(2\pi)^3\delta^3\left(\vec{k}_1+\vec{k}_2+\vec{k}_3\right)
\frac{H^4}{16\ep^2}\prod_{i=1}^3\frac{1}{k^3_i}
\frac{e^{\frac i2 \left(\vec{k}_1\wedge \vec{k}_2+
\vec{k}_2\wedge \vec{k}_3+\vec{k}_1\wedge \vec{k}_3\right)}}{K}
\left(k_1^2 k_2^2+\, \text{perm.}\right).
\ee
Here $K=k_1+k_2+k_3$. Now similar calculations can be done for the  second term in the three-point 
function (\ref{zetathree1}). Let us represent it as $(b)$,
\be
(b)=i \int_{t_0}^t dt^\pr \lan 0|{\cal H}(t^\pr)\zeta_\theta(x_1)\zeta_\theta(x_2)\zeta_\theta(x_3)|0
\ran.\label{b1}
\ee
The contribution due to this term in momentum space is given by 
(\ref{bapdsol})
\be
(b)=\ep(2\pi)^3\delta^3\left(\vec{k}_1+\vec{k}_2+\vec{k}_3\right)
\frac{H^4}{16\ep^2}\prod_{i=1}^3\frac{1}{k^3_i}
\frac{e^{\frac i2 \left(\vec{k}_1\wedge \vec{k}_2+
\vec{k}_2\wedge \vec{k}_3+\vec{k}_1\wedge \vec{k}_3\right)}}{K}
\left(k_1^2 k_2^2+\, \text{perm.}\right).\nonumber\\
\label{bsol}
\ee
So the contribution to the three-point function of $\zeta$ due 
to the first term of (\ref{zetathreepoint}) in Fourier space is given as
\bea
\lan\zeta_\theta(\vec{k}_1,t)\zeta_\theta(\vec{k}_2,t)
\zeta_\theta(\vec{k}_3,t)\ran
&=&2\ep(2\pi)^3\delta^3\left(\vec{k}_1+\vec{k}_2+\vec{k}_3\right)
\frac{H^4}{16\ep^2}\prod_{i=1}^3\frac{1}{k^3_i}\nonumber\\
&\times&\frac{e^{\frac i2 \left(\vec{k}_1\wedge \vec{k}_2+
\vec{k}_2\wedge \vec{k}_3+\vec{k}_1\wedge \vec{k}_3\right)}}{K}
\left(k_1^2 k_2^2+\, \text{perm.}\right).\nonumber\\
\label{three1}
\eea
This concludes the calculations of the three-point function of the redefined
field $\zeta_n$. Now to get the final three-point function of the field 
$\zeta$ we need to  consider the second and third term of equation 
(\ref{zetathreepoint}) coming due to field redefinitions. The contribution
to the three-point function due to first of these terms can be obtained using 
Wick's theorem and is given as
\be
\lan \zeta_\theta(x_1)\zeta_\theta(x_2)\zeta_\theta(x_3)=
\frac{\eta-\ep}{4}
\left(\lan\zeta_\theta(x_1)\zeta_\theta(x_2)\ran 
\lan\zeta_\theta(x_1)\zeta_\theta(x_3)\ran+\, \text{perm.} \right).
\ee
Now
\be
\lan\zeta_\theta(x_1)\zeta_\theta(x_2)\ran=\int \frac{d^3k_2}{(2\pi)^3}
\frac{H^2}{4\ep}\frac{1}{k^3_2}e^{-H\ora{\theta^0}\cdot\vec{k}_2}
e^{i\vec{k}_2\cdot\left(\vec{x}_1-\vec{x}_2\right)}.
\ee
So
\bea
\lan\zeta_\theta(x_1)\zeta_\theta(x_2)\ran
\lan\zeta_\theta(x_1)\zeta_\theta(x_3)\ran&=&
\int \frac{d^3k_2d^3k_3}{(2\pi)^9}
\frac{H^4}{16\ep^2}\frac{1}{k^3_2k_3^3}
e^{-H\ora{\theta^0}\cdot\left(\vec{k}_2+k_3\right)}
e^{i\left(\vec{k}_2+\vec{k}_3\right)\cdot\vec{x}_1-
i\vec{k}_2\cdot\vec{x}_2-i\vec{k}_3\cdot\vec{x}_3}\nonumber\\
&=&(2\pi)^3\int \frac{d^3k_1d^3k_2d^3k_3}{(2\pi)^9}
\delta^3\left(\vec{k}_1+\vec{k}_2+\vec{k}_3\right)\frac{H^4}{16\ep^2}
\frac{1}{k^3_2k_3^3}\nonumber\\
&\times& e^{H\ora{\theta^0}\cdot\vec{k}_1}
e^{-i\vec{k}_1\cdot\vec{x}_1-
i\vec{k}_2\cdot\vec{x}_2-i\vec{k}_3\cdot\vec{x}_3}.
\eea
Here in the second step we have introduced a $\delta$ function with integral 
over $k_1$ so that it matches with the results of the  rest of the terms. 
So the contribution to the 
three-point function due to first field redefinition term in momentum space
will be
\be
\lan\zeta_\theta(\vec{k}_1,t)\zeta_\theta(\vec{k}_2,t)
\zeta_\theta(\vec{k}_3,t)\ran
=\frac{\eta-\ep}{2}(2\pi)^3\delta^3\left(\vec{k}_1+\vec{k}_2+\vec{k}_3\right)
\frac{H^4}{16\ep^2}\prod_{i=1}^3\frac{1}{k^3_i}
\left(\sum_i k_i^3e^{H\ora{\theta^0}\cdot\vec{k}_i}\right).\label{threef2}
\ee
Now consider the second field redefinition term in Eq.~(\ref{zetathreepoint})
the contribution due to that is given as
\be
\lan \zeta_\theta(x_1)\zeta_\theta(x_2)\zeta_\theta(x_3)=
\frac{\ep}{2}\left(
\p_{x_1}^{-2}\left(\lan\zeta_\theta(x_1)\zeta_\theta(x_2)\ran
\p_{x_1}^{2}\lan\zeta_\theta(x_1)\zeta_\theta(x_3)\ran \right)+\, \text{perm.}\right).
\ee
Now 
\be
\p_{x_1}^{2}\lan\zeta_\theta(x_1)\zeta_\theta(x_3)\ran=
-\int \frac{d^3k_3}{(2\pi)^3}
\frac{H^2}{4\ep}\frac{1}{k_3}e^{-H\ora{\theta^0}\cdot\vec{k}_3}
e^{i\vec{k}_3\cdot\left(\vec{x}_1-\vec{x}_3\right)}.
\ee
So
\bea
\lan\zeta_\theta(x_1)\zeta_\theta(x_2)\ran
\p_{x_1}^{2}\lan\zeta_\theta(x_1)\zeta_\theta(x_3)\ran&=&
-\int \frac{d^3k_2d^3k_3}{(2\pi)^6}
\frac{H^4}{16\ep^2}\frac{1}{k^3_2k_3}
e^{-H\ora{\theta^0}\cdot\left(\vec{k}_2+k_3\right)}
e^{i\left(\vec{k}_2+\vec{k}_3\right)\cdot\vec{x}_1-
i\vec{k}_2\cdot\vec{x}_2-i\vec{k}_3\cdot\vec{x}_3}\nonumber\\
&=&-(2\pi)^3\int \frac{d^3k_1d^3k_2d^3k_3}{(2\pi)^9}
\delta^3\left(\vec{k}_1+\vec{k}_2+\vec{k}_3\right)\frac{H^4}{16\ep^2}
\frac{1}{k^3_2k_3}\nonumber\\
&\times& e^{H\ora{\theta^0}\cdot\vec{k}_1}
e^{-i\vec{k}_1\cdot\vec{x}_1-
i\vec{k}_2\cdot\vec{x}_2-i\vec{k}_3\cdot\vec{x}_3}.
\eea
So
\bea
\p_{x_1}^{-2}\left(\lan\zeta_\theta(x_1)\zeta_\theta(x_2)\ran
\p_{x_1}^{2}\lan\zeta_\theta(x_1)\zeta_\theta(x_3)\ran \right)&=&
(2\pi)^3\int \frac{d^3k_1d^3k_2d^3k_3}{(2\pi)^9}
\delta^3\left(\vec{k}_1+\vec{k}_2+\vec{k}_3\right)\frac{H^4}{16\ep^2}
\frac{1}{k_1^2k^3_2k_3}\nonumber\\
&\times& e^{H\ora{\theta^0}\cdot\vec{k}_1}
e^{-i\vec{k}_1\cdot\vec{x}_1-
i\vec{k}_2\cdot\vec{x}_2-i\vec{k}_3\cdot\vec{x}_3}.
\eea
So the contribution due to this term in Fourier space will be
\be
\lan\zeta_\theta(\vec{k}_1,t)\zeta_\theta(\vec{k}_2,t)
\zeta_\theta(\vec{k}_3,t)\ran
=\frac{\ep}{2}(2\pi)^3\delta^3\left(\vec{k}_1+\vec{k}_2+\vec{k}_3\right)
\frac{H^4}{16\ep^2}\prod_{i=1}^3\frac{1}{k^3_i}
\left(\sum_{i\not=j} k_ik_j^2e^{H\ora{\theta^0}\cdot\vec{k}_i}\right).
\label{threef3}
\ee
Now combining all the results from (\ref{three1}), (\ref{threef2}), 
(\ref{threef3}) for the various contributions to the three-point function of
$\zeta$, we get the final three-point function using 
Eq.~(\ref{zetathreepoint}) in momentum space as
\be
\lan\zeta_\theta(\vec{k}_1,t)\zeta_\theta(\vec{k}_2,t)
\zeta_\theta(\vec{k}_3,t)\ran
=(2\pi)^3\delta^3\left(\vec{k}_1+\vec{k}_2+\vec{k}_3\right)
\frac{H^4}{16\ep^2}\prod_{i=1}^3\frac{1}{k^3_i}{\cal A}\label{ztf}
\ee
where 
\bea
{\cal A}&=&4\ep\frac{e^{\frac i2 \left(\vec{k}_1\wedge \vec{k}_2+
\vec{k}_2\wedge \vec{k}_3+\vec{k}_1\wedge \vec{k}_3\right)}}{K}
\left(\sum_{i<j}k_i^2 k_j^2\right)+\frac{\eta-\ep}{2}
\left(\sum_i k_i^3e^{H\ora{\theta^0}\cdot\vec{k}_i}\right)\nonumber\\
&+&\frac{\ep}{2}\left(\sum_{i\not=j} k_ik_j^2e^{H\ora{\theta^0}\cdot\vec{k}_i}\right).
\label{threea}
\eea
This is the main result of this paper. In the limit 
$\theta^{\mu\nu}\rightarrow 0$ the above expression becomes similar to 
expression for the three-point function in commutative spacetime 
(Eq.~(4.5) and (4.6) of Maldacena \cite{Maldacena:2002vr}). 
 Now due to translational
invariance $\vec{k}_1+\vec{k}_2+\vec{k}_3=0$ and on comparing our results
with  the commutative case \cite{Maldacena:2002vr, Seery:2005wm} we see that 
the first term  in (\ref{threea})  is modified due to a phase 
factor that depends on $\theta_{ij}$, while the second and the last terms are 
modified by exponential factors. These modifications in the three-point function are due to the 
non-gaussian nature of noncommutativity. As also mentioned by \cite{Akofor:2007fv}, the $n$-point 
correlation functions for noncommutative fields are, in general, non-gaussian and cannot be expressed
as sums of products of two-point correlation function even in the absence of interactions.   
The three-point correlation function here is complex so to see its observational effects we again take  its self adjoint given as 
\cite{Koivisto:2010fk}
\bea
{\lan\zeta_\theta(\vec{k}_1,t)\zeta_\theta(\vec{k}_2,t)
\zeta_\theta(\vec{k}_3,t)\ran}_M&=&\frac12
\left(\lan\zeta_\theta(\vec{k}_1,t)\zeta_\theta(\vec{k}_2,t)
\zeta_\theta(\vec{k}_3,t)\ran+
\lan\zeta_\theta(-\vec{k}_1,t)\zeta_\theta(-\vec{k}_2,t)
\zeta_\theta(-\vec{k}_3,t)\ran\right)\nonumber\\
&=&(2\pi)^3\delta^3\left(\vec{k}_1+\vec{k}_2+\vec{k}_3\right)
\frac{H^4}{16\ep^2}\prod_{i=1}^3\frac{1}{k^3_i}
\left[\frac{4\ep\cos\left(\frac{\vec{k}_1\wedge\vec{k}_2}{2}\right)}{K}
\left(\sum_{i<j}k_i^2 k_j^2\right)\right.\nonumber\\
&+&\left. \frac{\eta-\ep}{2}
\left(\sum_i k_i^3\cosh\left(H\ora{\theta^0}\cdot\vec{k}_i\right)\right)
+\frac{\ep}{2}\left(\sum_{i\not=j} k_ik_j^2
\cosh\left(H\ora{\theta^0}\cdot\vec{k}_i\right)\right)\right].\nonumber\\
\label{asaf}
\eea
\section{Implications for observations}
\label{io}
The non-gaussianity in CMB is described in terms of the angular three-point
correlation functions in harmonic space called as "angular bispectrum", which
is related to the three-dimensional bispectrum of the primordial curvature
perturbations defined as \cite{Komatsu:2010hc,Dimastrogiovanni:2010sm}
\be
\lan\zeta(\vec{k}_1,t)\zeta(\vec{k}_2,t)\zeta(\vec{k}_3,t)\ran=
(2\pi)^3\delta^3\left(\vec{k}_1+\vec{k}_2+\vec{k}_3\right)
B_\zeta\left(k_1,k_2,k_3\right).\label{bid}
\ee 
We can generalize the above definition of bispectrum for the twisted quantum 
fields in noncommutative space time and it can be expressed using (\ref{asaf})
as 
\bea
B_{\zeta_\theta}\left(\vec{k}_1,\vec{k}_2,\vec{k}_3\right)&=&
\frac{H^4}{16\ep^2}\prod_{i=1}^3\frac{1}{k^3_i}\left[
\frac{4\ep\cos\left(\frac{\vec{k}_1\wedge\vec{k}_2}{2}\right)}{K}
\left(\sum_{i<j}k_i^2k_j^2\right)+\frac{\eta-\ep}{2}
\left(\sum_i k_i^3\cosh\left(H\ora{\theta^0}\cdot\vec{k}_i\right)\right)\
\right.\nonumber\\
&+&\left.\frac{\ep}{2}\left(\sum_{i\not=j} k_ik_j^2\cosh
\left(H\ora{\theta^0}\cdot\vec{k}_i\right)\right)\right].\label{bzeta}
\eea
Here the bispectrum also breaks the statistical isotropy. The anisotropic
bispectrum also arises in the cases where the vector fields are also present
during inflation \cite{Yokoyama:2008xw,Dimastrogiovanni:2010sm}. In 
\cite{Bartolo:2011ee} the method to analyze these models in the light of new CMB data is derived. 
Current observational limits on non-gaussianity are given in terms of a 
non-linearity parameter $f_{NL}$ that determines the amplitude and
scale dependence of non-gaussianity. We define $f_{NL}$ in a similar way as 
\cite{Yokoyama:2008xw, Dimastrogiovanni:2010sm} where it is assumed that the corrections to 
the standard power spectrum due to statistical anisotropy are very small. So
\be
f_{NL}=\frac56
\frac{B_{\zeta_\theta}\left(\vec{k}_1,\vec{k}_2,\vec{k}_3\right)}
{P_{\zeta}(k_1)P_{\zeta}(k_2)+
P_{\zeta}(k_2)P_{\zeta}(k_3)
+P_{\zeta}(k_1)P_{\zeta}(k_3)}.
\ee
Using the power spectrum (\ref{powers}) it becomes
\bea
f_{NL}&=&\frac56\frac{1}{\sum_i k_i^3}\left[
4\ep\frac{\cos{\left(\frac{\vec{k}_1\wedge \vec{k}_2}{2}\right)}}{K}
\sum_{i<j}\left(k_i^2 k_j^2\right)+\frac{\eta-\ep}{2}
\left(\sum_i k_i^3\cosh\left(H\ora{\theta^0}\cdot\vec{k}_i\right)\right)\right.
\nonumber\\
&+&\left.\frac{\ep}{2}\left(\sum_{i\not=j} k_ik_j^2\cosh
\left(H\ora{\theta^0}\cdot\vec{k}_i\right)\right)\right].
\eea
This kind of $f_{NL}$ generally arises where  the curvature 
perturbation is expressed as  $\zeta_g=\zeta_g+\frac35\zeta_g^2$ and 
$f_{NL}$ peaks at the so called squeezed triangle limit defined as
 $|\vec{k}_1|=|\vec{k}_2|=k$ and $|\vec{k_3}|<<k$. So in a similar fashion
the $f_{NL}$ for noncommutative case in the above limit is given as 
\be
f_{NL}=\frac{5}{12} 
\left[2\ep\cos{\left(\frac{\vec{k}_1\wedge \vec{k}_2}{2}\right)}
+\frac{\eta}{2}\left(\cosh\left(H\ora{\theta^0}\cdot\vec{k}_1\right)+
\cosh\left(H\ora{\theta^0}\cdot\vec{k}_2\right)\right)\right].
\label{fnlsq}
\ee
It is clear from the above expression that the amplitude of
 $f_{NL}$  is very small and of the order of slow-roll parameters
for the case of small statistical anisotropy. But it has
 scale dependence and direction dependence that can help us to distinguish it 
from the commutative case. 
The current limits on the amplitude of $f_{NL}$ for squeezed triangle limit 
 are $f_{NL}=2.7\pm 5.8$ from the recently released
PLANCK data \cite{Ade:2013ydc} and  
$f_{Nl}=48\pm 20$ from large scale structure probes
\cite{Xia:2011hj} at $68\%$ confidence level. One can define the scale dependence of $f_{NL}$ by
a parameter $n_{NG}$ analogous to the spectral index \cite{Chen:2006nt}
\be
n_{NG}=\frac{d \ln|f_{NL}|}{d\ln k}.
\label{nng}
\ee 
To quantify the scale dependence coming due to noncommutativity, we assume 
$\hat{\theta^0}$  along $\vec{k_1}$ and hence $n_{NG}$ due to first  term of 
Eq.~(\ref{fnlsq}) (term depending on $\theta_{ij}$) can be obtained as
\be
n_{NG}=-k_1^i\theta_{ij} k_2^j\tan\left(\frac{k_1^i\theta_{ij} k_2^j}{2}\right).
\ee
 And similarly for the second term of Eq.~(\ref{fnlsq}), terms depending on $\theta^0$, $n_{NG}$ is given as
\be
n_{NG}=H\theta^0 k \tanh(H\theta^0 k).
\ee 
 The running of the non-gaussianity $n_{NG}$ 
for $\ora{\theta^0}=0$ in our case is similar to \cite{Koivisto:2010fk} (their $n_{f_{NL}}=n_{NG}$) and
they argued that the detection of $n_{NG}$ could put strong bounds on $\theta_{ij}$.
The constraints on the running of the non-gaussianity with ongoing and future large scale structure
surveys and CMB observations were studied  in \cite{Sefusatti:2009xu, Becker:2012yr} 
and they showed that we will be  able to constraint 
$n_{NG}$ with a $1-\sigma$ uncertainty of $\Delta n_{NG}\sim0.1$. 
Taking into account the bounds on noncommutativity scale $H\theta^0<0.01$ claimed by Akofor et al. 
\cite{Akofor:2008gv}, the running of non-gaussianity arising due to the term depending on $\theta^0$ is of the
order of $10^{-7}$ for the pivot scale $k=0.05$MPc\textsuperscript{-1} which is far beyond the current 
reachable limit.  
Since the amplitude of $f_{NL}$ with the noncommutative geometry is of the 
order of slow-roll parameters, the scale  dependence of $f_{Nl}$ due to noncommutativity 
with ongoing and planned  observations of CMB and LSS is undetectable.
\section{Conclusions}
\label{cnc}
Detection of primordial non-gaussianity in the CMB anisotropy  and large
scale structure is the main challenge of current and future observations and it
can play an important role in discriminating various models of 
inflation. 
In this paper we have calculated the primordial non-gaussianity in single field
inflation with spacetime noncommutativity. We have used Maldacena's approach
\cite{Maldacena:2002vr} to compute the two-point and three-point correlation 
functions for the comoving curvature perturbation $\zeta$ for the 
noncommutative case described by \cite{Akofor:2007fv}. 
Both the power spectrum and the bispectrum for this model are direction 
dependent and breaks the statistical isotropy due to the preferred direction 
of $\hat\theta$. This direction dependent power spectrum  was analyzed by
\cite{Akofor:2008gv} to put constraints on the scale of noncommutativity in 
the light of  WMAP5, ACBAR and CBI data and it was concluded  that the WMAP5 data at high $l$
is not sufficient to constraint the noncommutative scale $\theta$ and using one-parameter 
$\chi^2$ analysis they claimed that $H\theta^0<0.01$MPc. Since
recently released PLANCK data gives the  CMB temperature anisotropy power spectra up to $l\ge 2500$
with better precision, the author and collaborators plan to  analyze the power 
spectrum (\ref{powernc}) with the  PLANCK  and other LSS data. The breaking of statistical isotropy
detected by PLANCK i.e dipolar modulation and hemispherical power asymmetry can not be explained with 
the power spectrum (\ref{powernc}) as it is parity conserving. But with some modifications, as 
in \cite{Koivisto:2010fk}, the hemispherical power asymmetry can be generated with noncommutative 
spacetime \cite{Koivisto:2010fk}.
   
The statistical anisotropic bispectrum can be extracted from 
the three-point correlation function of CMB \cite{Bartolo:2011ee} and for 
$f_{NL}\approx 30$, future experiments could be sensitive to a ratio of the 
anisotropic to the isotropic amplitudes of the bispectrum up to $10\%$.
The amplitude of the non-linearity parameter $f_{NL}$ for our case is very 
small for small statistical anisotropy but it has a  scale dependence different then commutative case. 
Ongoing PLANCK and future CMB and large scale 
structure observations would be able to measure the running of 
non-gaussianity up-to $1-\sigma$ uncertainty of $\Delta n_{NG}\sim0.1$ 
\cite{Becker:2012yr}. Since the effects on the scale dependence of $f_{NL}$ 
due to noncommutativity are very small, 
it is difficult to distinguish these effects from the commutative case in the light of current 
observations.  
 
\acknowledgments
We thank  A. P. Balachandran for very helpful discussions and 
valuable suggestions.
 
\appendix
\section{Some calculation details}
The integral appearing in the calculation of the first term in 
the three-point function (\ref{zetathree1}) in Fourier space can be read from 
Eq.~(\ref{a2}) as 
\bea
(a)&=&-4i\ep^2\int dt^\pr a^5 H \int d^3x\int\prod_{i=4}^6\frac{d^3k_i}{k^2_6(2\pi)^9}
\lan 0|\zeta(\vec{k}_1,t_1)\zeta(\vec{k}_2,t_2)\zeta(\vec{k}_3,t_3)
\dot\zeta(\vec{k}_4,t_4)\dot\zeta(\vec{k}_5,t_5)\dot\zeta(\vec{k}_6,t_6)|0\ran\nonumber\\
&\times&e^{i\left(\vec{k}_4\cdot\vec{x}+\vec{k}_5\cdot\vec{x}+\vec{k}_6\cdot\vec{x}\right)}
e^{\frac i2 {\cal P}},\label{aapd}
\eea
where $t_i$s are given by Eq.~(\ref{time}).
Now we will calculate the six-point function entering in the above integrand
 separately and denote it  as $A$. So
\be
A=\lan 0|\zeta(\vec{k}_1,t_1)\zeta(\vec{k}_2,t_2)\zeta(\vec{k}_3,t_3)
\dot\zeta(\vec{k}_4,t_4)\dot\zeta(\vec{k}_5,t_5)\dot\zeta(\vec{k}_6,t_6)|0\ran.
\label{Aa}
\ee
Now using Wick's theorem and leaving the disconnected diagrams we will get 6 terms in above
expression. Let us consider one of them and denote it by $A_1$ so
\be
A_1= \lan 0|\left[\zeta^+(\vec{k}_1,t_1),\dot\zeta^-(\vec{k}_4,t_4)\right]
\left[\zeta^+(\vec{k}_2,t_2),\dot\zeta^-(\vec{k}_5,t_5)\right]
\left[\zeta^+(\vec{k}_3,t_3),\dot\zeta^-(\vec{k}_6,t_6)\right]
|0\ran
\ee
where the $\zeta^+$ and $\zeta^-$ denote the positive and negative frequency 
part of the quantum field $\zeta$ (see (\ref{zetafourier})).
Now since
\be
\left[\zeta^+(\vec{k}_1,t_1),\dot\zeta^-(\vec{k}_4,t_4)\right]=(2\pi)^3
\delta^3\left(\vec{k}_1+\vec{k}_4\right)u\left(\vec{k}_1,t_1\right)
\dot u^\star\left(-\vec{k}_4,t_4\right),
\ee
we have
\bea
A_1&=&(2\pi)^9\delta^3\left(\vec{k}_1+\vec{k}_4\right)\delta^3\left(\vec{k}_2+\vec{k}_5\right)
\delta^3\left(\vec{k}_3+\vec{k}_6\right)u\left(\vec{k}_1,t_1\right) u\left(\vec{k}_2,t_2\right)
u\left(\vec{k}_3,t_3\right)\nonumber\\&\times&\dot u^\star\left(-\vec{k}_4,t_4\right)
\dot u^\star\left(-\vec{k}_5,t_5\right)\dot u^\star\left(-\vec{k}_6,t_6\right).
\eea
Putting this back to the integral (\ref{a1}) and denoting the contribution due to this term as
$(a)_1$ and doing the delta integrals we get
\bea
(a)_1&=&-4i\ep^2\int_{t_0}^t dt^\pr a^5 H \int d^3x\frac{1}{k^2_3}
u\left(\vec{k}_1,t_1\right) u\left(\vec{k}_2,t_2\right)u\left(\vec{k}_3,t_3\right)
\dot u^\star\left(\vec{k}_1,t_4\right)
\dot u^\star\left(\vec{k}_2,t_5\right)\dot u^\star\left(\vec{k}_3,t_6\right)\nonumber\\
&\times& e^{i\left(\vec{k}_4\cdot\vec{x}+\vec{k}_5\cdot\vec{x}+\vec{k}_6\cdot\vec{x}\right)}
e^{\frac i2 {\cal P}_1}\nonumber\\
&=&-4i\ep^2(2\pi)^3\delta^3\left(\vec{k}_1+\vec{k_2}+\vec{k_3}\right)\int_{t_0}^t
dt^\pr a^5 H\frac{1}{k^2_3}u\left(\vec{k}_1,t_1\right) u\left(\vec{k}_2,t_2\right)u\left(\vec{k}_3,t_3\right)\nonumber\\
&\times&\dot u^\star\left(\vec{k}_1,t_4\right)
\dot u^\star\left(\vec{k}_2,t_5\right)\dot u^\star\left(\vec{k}_3,t_6\right)e^{\frac i2 {\cal P}_1}\label{a3}
\eea
where ${\cal P}_1={\left.{\cal P}\right|}_{\vec{k}_4=-\vec{k}_1,\, 
\vec{k}_5=-\vec{k}_2,\, \vec{k}_6=-\vec{k}_3}$ and $t_i$s are also
calculated using these values of momenta.
Here the limit of integration goes from $t_0=-\infty$ to $t=\infty$.
To solve the
integral we will go to conformal time where the above limits correspond to
$\tau\rightarrow (-\infty,0)$. Now from Eq.~(\ref{uksol}) we have
\be
u(\vec{k},\tau)=\frac{v_{\vec{k}}}{z}=\frac{i H}{\sqrt{4\epsilon k^3}}
\left(1+ik\tau\right)e^{-ik\tau}.\label{uksol1}
\ee
Since from Eq.~(\ref{taut}) we know that the conformal time corresponding
to $t_i$s will be $\tau\times e^{\theta\, \text{dependent term} }$ and from
Eq.~(\ref{time}) we have terms like $(t+\theta\, \text{dependent term})$ for
$t_1,\, t_2,\, t_3$ so for $t\rightarrow\infty$ or $\tau\rightarrow 0$
conformal time corresponding to $t_1,\, t_2, \, t_3$ will be zero.
So in conformal time $u\left(\vec{k}_1,t_1\right) 
u\left(\vec{k}_2,t_2\right)u\left(\vec{k}_3,t_3\right) \rightarrow 
u\left(\vec{k}_1,0\right) u\left(\vec{k}_2,0\right)u\left(\vec{k}_3,0\right)$.
Now we denote the conformal time corresponding to $t^\pr$ by $\tau$ so
$\dot u^\star\left(\vec{k}_1,t_4\right)\rightarrow 
\frac{1}{a}\frac {du^\star\left(\vec{k}_1,\tau_4\right)}{d\tau}$ and from
(\ref{taut}) and (\ref{time}) we get
$\tau_4=\tau_5=\tau_6=\tau e^\frac{H\ora{\theta^0}\cdot\left(\vec{k}_1
+\vec{k}_2+\vec{k}_3\right)}{2}$.
Now
\be
\frac {du^\star\left(\vec{k}_1,\tau_4\right)}{d\tau}=
\frac{-i H}{\sqrt{4\epsilon k^3}}k_1^2\tau
e^{H\ora{\theta^0}\cdot\left(\vec{k}_1+\vec{k}_2+\vec{k}_3\right)}
e^{ik_1\tau e^\frac{H\ora{\theta^0}\cdot\left(\vec{k}_1
+\vec{k}_2+\vec{k}_3\right)}{2}}.
\ee
Now due to translational invariance of de Sitter space $\vec{k}_1+\vec{k}_2+\vec{k}_3=0$. So, the integral (\ref{a3}) becomes
\bea
(a)_1&=&-i\ep(2\pi)^3\delta^3\left(\vec{k}_1+\vec{k}_2+\vec{k}_3\right)
\frac{H^7}{16\ep^2}\prod_{i=1}^3\frac{1}{k^3_i}\nonumber\\
&\times&\int_{-\infty}^0 a^3\tau^3 k_1^2 k^2_2e^{\frac i2 {\cal P}_1}
e^{iK\tau}\nonumber\\ 
&=&\ep(2\pi)^3\delta^3\left(\vec{k}_1+\vec{k}_2+\vec{k}_3\right)
\frac{H^4}{16\ep^2}\prod_{i=1}^3\frac{1}{k^3_i} \frac{k_1^2 k_2^2}{K}
e^{\frac i2 {\cal P}_1}e^{\frac{5H\ora{\theta^0}\cdot\left(\vec{k}_1
+\vec{k}_2+\vec{k}_3\right)}{2}}.
\eea
Here we have rotated the contour from $(-\infty,\, 0)$ to $i(\infty,\, 0)$
and $K=k_1+k_2+k_3$. Now to calculate ${\cal P}_1$ let us recall
(\ref{phase})
\be
{\cal P}=\vec{k}_1\wedge \vec{k}_2+
\vec{k}_2\wedge \vec{k}_3+\vec{k}_1\wedge \vec{k}_3
+\vec{k}_1\wedge\left(\vec{k}_4+\vec{k}_5+\vec{k}_6\right)
+\vec{k}_2\wedge\left(\vec{k}_4+\vec{k}_5+\vec{k}_6\right)+
\vec{k}_3\wedge\left(\vec{k}_4+\vec{k}_5+\vec{k}_6\right).
\ee
Hence ${\cal P}_1=\vec{k}_1\wedge \vec{k}_2+
\vec{k}_2\wedge \vec{k}_3+\vec{k}_1\wedge \vec{k}_3$.
Now rest of the terms in (\ref{Aa}) can be found be different permutations of
$k_4,\, k_5,\, k_6 $ and the phase factors will be same as ${\cal P}_1$ after
imposing the different conditions due to delta function integrals. So from
equation (\ref{a2}) we get the first term of the right hand
side of equation (\ref{zetathree1}) in Fourier space as
\be
(a)=\ep(2\pi)^3\delta^3\left(\vec{k}_1+\vec{k}_2+\vec{k}_3\right)
\frac{H^4}{16\ep^2}\prod_{i=1}^3\frac{1}{k^3_i}
\frac{e^{\frac i2 \left(\vec{k}_1\wedge \vec{k}_2+
\vec{k}_2\wedge \vec{k}_3+\vec{k}_1\wedge \vec{k}_3\right)}}{K}
\left(k_1^2 k_2^2+\, \text{perm.}\right).\label{aapdsol}
\ee
Now  the second term in the three-point function (\ref{zetathree1})
 is denoted as $(b)$ and can be read from Eq.~(\ref{b1}) as
\bea
(b)&=&i \int_{t_0}^t dt^\pr \lan 0|{\cal H}(t^\pr)\zeta_\theta(x_1)\zeta_\theta(x_2)\zeta_\theta(x_3)|0\ran\nonumber\\
  &=&-4 i\ep^2\int dt^\pr a^5 H \int d^3x
\lan 0|{\left.\dot\zeta^2\p^{-2}\dot\zeta\right|}_{t^\pr,\vec{x}}
e^{\frac12\ola{\p_{x_\mu}}\wedge P_\nu}
\zeta_\theta(x_1)\zeta_\theta(x_2)\zeta_\theta(x_3)|0\ran\nonumber\\
&=&-4 i\ep^2\int dt^\pr a^5 H \int d^3x
\lan 0|{\left.\dot\zeta^2\p^{-2}\dot\zeta\right|}_{t^\pr,\vec{x}}
e^{-\frac i2\ola{\p_{x}}\wedge \left(\ora{\p_{x_1}}+\ora{\p_{x_2}}+
\ora{\p_{x_3}}\right)}\nonumber\\
&\times&\zeta(x_1)\zeta(x_2)\zeta(x_3)|0\ran
e^{-\frac i2\left(\ola{\p_{x_1}}\wedge \ola{\p_{x_2}}+
\ola{\p_{x_2}}\wedge \ola{\p_{x_3}}+\ola{\p_{x_1}}\wedge \ola{\p_{x_3}}\right)}.
\eea
Here we have used (\ref{zetatheta}). Now in the Fourier space we get
\bea
(b)&=&-4 i\ep^2\int dt^\pr a^5 H \int d^3x\int \prod_{i=1}^6 \frac{d^3k}
{-k_6^2(2\pi)^{18}}e^{i\left(\vec{k}_4+\vec{k}_5+\vec{k}_6\right)\cdot \vec{x}}
e^{-\frac i2\ola{\p_{x}}\wedge \left(\ora{\p_{x_1}}+\ora{\p_{x_2}}+
\ora{\p_{x_3}}\right)}\nonumber\\
&\times&\lan 0|\dot\zeta(\vec{k}_4,t^\pr)\dot\zeta(\vec{k}_5,t^\pr)
\dot\zeta(\vec{k}_6,t^\pr)
\zeta(\vec{k}_1,t_1)\zeta(\vec{k}_2,t_2)\zeta(\vec{k}_3,t_3)|0\ran\nonumber\\
&\times&e^{i\left(\vec{k}_1\cdot\vec{x}_1+\vec{k}_2\cdot\vec{x}_2+
\vec{k}_3\cdot\vec{x}_3\right)}
e^{-\frac i2\left(\ola{\p_{x_1}}\wedge \ola{\p_{x_2}}+
\ola{\p_{x_2}}\wedge \ola{\p_{x_3}}+\ola{\p_{x_1}}\wedge \ola{\p_{x_3}}\right)}
\nonumber\\
&=&4 i\ep^2\int dt^\pr a^5 H \int d^3x\int \prod_{i=1}^6 \frac{d^3k}
{k_6^2(2\pi)^{18}}e^{i\left(\vec{k}_4+\vec{k}_5+\vec{k}_6\right)\cdot \vec{x}}
e^{i\left(\vec{k}_1\cdot\vec{x}_1+\vec{k}_2\cdot\vec{x}_2+
\vec{k}_3\cdot\vec{x}_3\right)}\nonumber\\
&\times&\lan 0|\dot\zeta(\vec{k}_4,t_4)\dot\zeta(\vec{k}_5,t_5)
\dot\zeta(\vec{k}_6,t_6)
\zeta(\vec{k}_1,t_1)\zeta(\vec{k}_2,t_2)\zeta(\vec{k}_3,t_3)|0\ran
e^{\frac i2 \tilde{\cal P}}.\label{b1}
\eea
Here
\bea
\tilde{\cal P}&=&\vec{k}_1\wedge \vec{k}_2+\vec{k}_2\wedge \vec{k}_3
+\vec{k}_1\wedge \vec{k}_3-\left(\vec{k}_1+\vec{k}_2+\vec{k}_3\right)
\wedge\left(\vec{k}_4+\vec{k}_5+\vec{k}_6\right),\label{phaseb}\\
t_1&=&t+\frac{\ora{\theta^0}\cdot
\left(\vec{k}_2+\vec{k}_3-\vec{k}_4-\vec{k}_5-\vec{k}_6\right)}{2},\nonumber\\
t_2&=&t+\frac{\ora{\theta^0}\cdot
\left(-\vec{k}_1+\vec{k}_3-\vec{k}_4-\vec{k}_5-\vec{k}_6\right)}{2},\nonumber\\
t_3&=&t+\frac{\ora{\theta^0}\cdot
\left(\vec{k}_1-\vec{k}_2-\vec{k}_4-\vec{k}_5-\vec{k}_6\right)}{2},\nonumber\\
t_4&=&t^\pr+\frac{\ora{\theta^0}\cdot\left(\vec{k}_1+\vec{k}_2+
\vec{k}_3\right)}{2},\nonumber\\
t_5&=&t^\pr+\frac{\ora{\theta^0}\cdot\left(\vec{k}_1+\vec{k}_2+
\vec{k}_3\right)}{2},\nonumber\\
t_6&=&t^\pr+\frac{\ora{\theta^0}\cdot\left(\vec{k}_1+\vec{k}_2+
\vec{k}_3\right)}{2}.\label{timeb}
\eea
Now all the calculations can be done for $(b)$ as earlier and the final answer
is
\be
(b)=\ep(2\pi)^3\delta^3\left(\vec{k}_1+\vec{k}_2+\vec{k}_3\right)
\frac{H^4}{16\ep^2}\prod_{i=1}^3\frac{1}{k^3_i}
\frac{e^{\frac i2 \left(\vec{k}_1\wedge \vec{k}_2+
\vec{k}_2\wedge \vec{k}_3+\vec{k}_1\wedge \vec{k}_3\right)}}{K}
\left(k_1^2 k_2^2+\, \text{perm.}\right).\label{bapdsol}
\ee

\end{document}